\documentclass[]{emulateapj}
\usepackage{graphicx}
\usepackage{amssymb}
\usepackage{amsmath}

\newcommand{\HI}{H{\small I }}

\begin{document}
\title{A technique for weak lensing with velocity maps: eliminating ellipticity noise in \HI radio observations}
\author{Miguel F. Morales\altaffilmark{1}}

\altaffiltext{1}{Harvard-Smithsonian Center for Astrophysics, 60 Garden Street, MS-51 P-225, Cambridge MA 02138-1516}

\begin{abstract}
Weak lensing surveys have become a powerful tool for mapping mass distributions and constraining the expansion history of our Universe, but continuum surveys must average over a large number of galaxies to average down the ellipticity noise due to the unknown ellipticity and orientation of the lensed galaxies. This Letter presents a technique for measuring weak lensing with velocity maps that avoids the ellipticity noise. By studying the inherent noise characteristics, we argue this new technique could rival or exceed the sensitivity of traditional continuum observations for upcoming \HI radio surveys.
\end{abstract}


\section{Introduction}

Weak lensing has become one of the most useful techniques for measuring foreground mass distributions, and more recently has emerged as one of the most promising ways of constraining the expansion history of the universe \citep{Miralda-Escude1991, KaiserWL,DETF}.

Weak lensing measurements traditionally work by measuring the increase in ellipticity and tangential alignment of background galaxies \citep{Miralda-Escude1991,BartelmannSchneiderWL}. Because the intrinsic ellipticity and alignment of each of the background galaxies is unknown, a large number of these galaxies must be averaged over. This effect is commonly called ``ellipticity noise,'' and requires weak lensing measurements to have a very high density of background galaxies.

This Letter presents a technique for measuring weak lensing with galactic velocity maps. The use of the projected velocity distribution allows weak lensing measurements based on a single galaxy, and avoids the ellipticity noise inherent in broad band lensing surveys. The same underlying effect has been previously identified by \citet{Blain2002}. In this Letter we clarify the underlying physical process, how it can be observed, and identify the error characteristics. The attributes of the observational errors are then used to argue that this technique could be significant for upcoming \HI observations.

Section \ref{VMsec} introduces how lensing can be identified in velocity maps, and the inherent error characteristics. Section \ref{Impsec} then analyzes the sensitivities, and argues that the velocity map technique may be of comparable sensitivity to broad band observations for upcoming \HI radio surveys.  Whether broad band or velocity map techniques are more sensitive is likely to depend on the details of the observation and instrument, and the currently unknown systematics.

\section{Lensed galactic velocity maps}
\label{VMsec}

In standard week lensing measurements the distribution of light of the observed galaxy is symmetric around the axis of rotation (in practice and by assumption). However, this is not true in velocity maps. The projection of the velocity provides a well-defined azimuthal structure in the galactic image. It is the azimuthal structure in the velocity map and how it changes under shear which allows us to measure the lensing signal and eliminate ellipticity noise.

\begin{figure*}
\begin{center}
\plotone{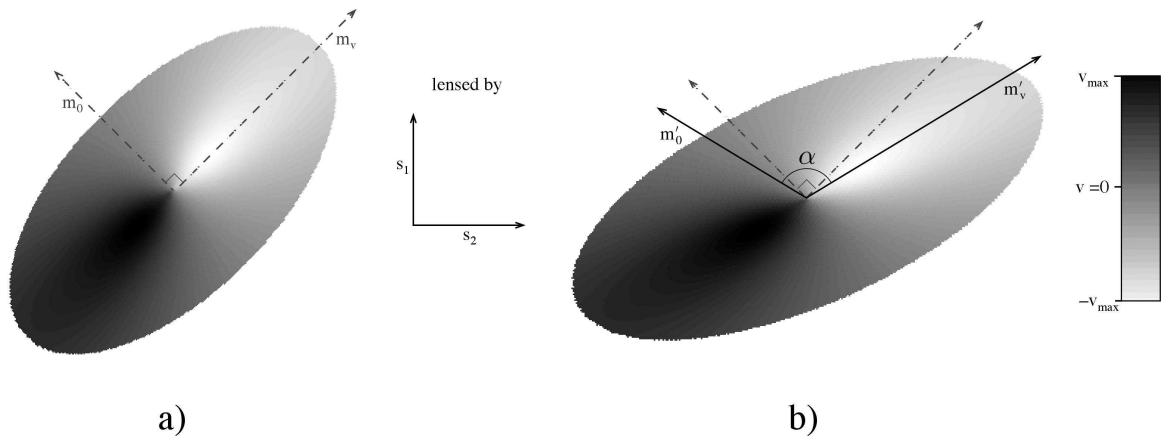}
\caption{This cartoon illustrates the effects of lensing on a velocity map. Panel a) shows a representative velocity map without lensing. The axis of rotation is oriented $60^\circ$ from the line of sight, with the grey value indicating the observed radial velocity. The image is then lensed by $s_1$ \& $s_2$ to produce the image in panel b). The orientation of the axis along the radial velocity maximum ($m_v$) and the zero velocity axis ($m_0$) are at right angles in the unlensed image, but have a larger angle $\alpha$ in the lensed image. The observed angle $\alpha$ directly measures the weak lensing, as described in detail in the text. In the limit of a shear-only field, $\alpha$ directly measures the component of the shear indicated by $s_1$ \& $s_2$, i.e. the shear component aligned with the bisector of $\alpha$. (For this example a large value of $\kappa = \gamma=0.2$, equivalent to an isothermal mass distribution centered directly below the lensed galaxy, was used for illustration purposes.)}
\label{lensfig}
\end{center}
\end{figure*}

In the absence of lensing, a simple disk galaxy will produce a cross-like velocity image as shown in panel a) of Figure \ref{lensfig}. For circular orbits the projected velocity is simply $\frac{v(r) m_v}{r}\sin \phi$, where $r$ is the distance from the center of the galaxy, $m_v$ is the projected distance along the major axis, and $\phi$ is the angle between the line of sight and the galaxy's rotation axis. This produces a maximum in velocity along the $m_v$ axis and a zero along the $m_0$ axis, which are necessarily separated by $90^\circ$ in the unlensed velocity map. This simple cross-like structure is what would be seen in the absence of a lensing source, or equivalently, is the structure seen at the plane of the lens.

To determine the lensed velocity map, we choose axes $s_1,s_2$ aligned at $45^\circ$ degrees to the axes of the initial galaxy. Following \citet{Miralda-Escude1991}, the image is then stretched by 
\begin{equation}
\label{ }
[1-(\kappa-\gamma)]^{-1}
\end{equation}
along the $s_1$ axis, and
\begin{equation}
\label{ }
[1-(\kappa+\gamma)]^{-1}
\end{equation}
along the $s_2$ axis, where $\kappa$ is the convergence and $\gamma$ is the shear along $s_1,s_2$. The resulting image is shown in panel b) of Figure \ref{lensfig}. Since the vertical ($s_1$) and horizontal ($s_2$) components of the unit vectors $m_0$ and $m_v$ are scaled by differing amounts, the  zero and maximum velocity axes in the lensed image ($m'_0, m'_v$) are no longer orthogonal.

The angle between the velocity axes in the lensed image is given by
\begin{equation}
\label{alphaeq}
\alpha = 2 \arctan\! \left[ \frac{(1-\kappa) + \gamma}{(1-\kappa)-\gamma} \right] .
\end{equation}
Measuring the angle between the velocity axes in the lensed image directly probes the weak lensing parameters.

The image in panel b) clearly shows the increase in ellipticity and tangential alignment of the lensed image that are used in standard weak lensing measurements. Note however, that while the zero and maximum velocity axes are aligned with the major and minor axes of the unlensed image, this is no longer true for the lensed image. 

To explore the properties of this distortion further, we note that $\alpha$ only depends on the convergence $\kappa$ to second order, and we can safely neglect this small contribution for the remainder of this discussion. To first order, $\alpha$ then only depends on the component of the shear orthogonal to the initial velocity axes $m_0,m_v$, or equivalently the component of the shear aligned with the bisector of $\alpha$ and shown by $s_1,s_2$ in Figure \ref{lensfig}. (Note that shear axes are orthogonal under $45^\circ$ rotations, see \citet{KamshearNotes2005} for a nice discussion of the mathematical properties.) shear along the orthogonal $m_0,m_v$ orientation will change the ellipticity of the observed image, but will not change the orientation of the velocity axes. To first order $\alpha$ measures the component of the shear field aligned with the bisector of $\alpha$.

The orientation of the shear component measured by $\alpha$ is determined by the observed orientation of the lensed galaxy. To measure the total shear in a small region thus necessitates determining $\gamma$ for two nearby galaxies with different observed orientations $\theta_1, \theta_2$. If we indicate the true shear field as $\epsilon_+,\epsilon_\times$ (often chosen so $\epsilon_\times = 0$), then the component of the shear measured with each galaxy is
\begin{eqnarray}
\gamma_1 = \epsilon_+ \cos 2 \theta_1 + \epsilon_\times \sin 2 \theta_1, \\
\gamma_2 = \epsilon_+ \cos 2 \theta_2 + \epsilon_\times \sin 2 \theta_2.
\end{eqnarray}
These can be easily solved to find 
\begin{equation}
\label{ }
\epsilon_+ = \frac{\gamma_2 \sin(2 \theta_1) - \gamma_1 \sin(2 \theta_2)}{\sin(2\theta_1-2\theta_2)}.
\end{equation}
If the orientation of the shear is already given by a mass model, then only one galaxy is needed to obtain a measurement.

An equivalent way of viewing the lensing transformation is that the lens sees a flat projected image of the galaxy with the characteristic cross-shape in the velocity map ($m_0 \perp m_v$ as shown in the left-hand panel). The observer then sees that image rotated around the $s_2$ axis (out of the image plane) by the lens. As the rotation around $s_2$ increases, the angle $\alpha$ also increases. Looking at the lens transformation as a rotation of the initial velocity map out of the image plane naturally leads to efficient algorithms for detecting the weak lensing signature in the observed velocity map.

\subsection{Error characteristics}
\label{ErrorSec}
The observational applicability of measuring weak lensing with velocity maps depends on the error characteristics inherent to the technique. The three primary error characteristics of velocity map lensing measurements are:

\emph{Reduced number counts.} The most obvious change is the reduced number counts. Because the ellipticity noise has been eliminated, in principal only two galaxies are needed to measure the shear as opposed to hundreds. This advantage is offset by the much smaller flux from emission lines compared to the integrated luminosity. The observational implications come down to the relative strength of the line emission, and the galaxy luminosity function, and is discussed in more depth in Section \ref{Impsec}.

\emph{Uncertainty depends on brightness of the image.} More importantly, the precision of the lensing measurement increases with the brightness of the galactic velocity image until limited by systematic errors. In standard weak lensing observations, bright galaxies are no more useful than faint galaxies due to the unknown intrinsic ellipticity, and all galaxies are weighted the same. With velocity maps, the uncertainty in the shear keeps decreasing with the uncertainty in the image. This makes bright galaxies much more valuable, and changes the way one tries to perform the observations (a few bright galaxies may be sufficient).

\emph{Reduced systematics.} Systematic errors can be more important than raw sensitivity, and velocity map lensing observations partially avoid three of the common systematics affecting continuum shear measurements. First, the velocity distribution is expected to be more circular than the distribution of light, and velocity lensing measurements are unaffected by bright knots in the galactic image. Second, the velocity map technique is insensitive to intrinsic alignments of the lensed galaxies. And finally, small errors in the telescope PSF become a second order effect instead of first order, and should be slightly less problematic for the velocity map technique.

These error characteristics are encouraging, and in the next section we discuss the feasibility of measuring weak lensing with velocity maps.

\section{Observational Feasibility}
\label{Impsec}

Because emission lines are much fainter than the integrated luminosity, any survey targeting galactic velocity maps will necessarily have a lower density of background galaxies than an equivalent broadband survey. However, as we showed in Section \ref{VMsec} the velocity map survey requires fewer galaxies due to the absence of ellipticity noise. In the end, which survey is more sensitive depends on the line-to-continuum luminosity ratio, the luminosity function ($dN/dS$) of the lensed galaxies, and the ability of the velocity method to identify the shear signal in low signal-to-noise images. 

Radio \HI observations are the most promising location in the electro-magnetic spectrum for producing sensitive galaxy surveys with resolved velocity maps. The \HI line is bright, the luminosity characteristics are relatively well understood, and several upcoming instruments will start taking much more sensitive \HI observations over the next few years.

The following sections review the relevant observational considerations and the sensitivity an \HI lensing survey might attain.

\subsection{Luminosity functions and SKA galaxy counts}

\citet{AbdallaRawlingsSKA} and \citet{BlakeSKA} have performed extensive studies of the continuum and \HI radio luminosity functions, and the resulting implications for radio cosmology with the SKA. Much like the optical galaxy luminosity function, the \HI luminosity function is a power-law (-1.3) with a knee at $M^*_{\rm \HI}$. For surveys above the knee, increased sensitivity quickly leads to higher number counts. Past the knee, however, enhancing the sensitivity only slowly increases the number counts.
 (See \citet{ZwaanHIPASS} for measurements of the \HI luminosity function at low redshift.)

\citet{BlakeSKA} have explored these consequences for proposed SKA galaxy surveys. For our purposes the most important result is that the number density of galaxies is $\gtrsim10$ arcmin$^{-2}$ in the \HI survey and $\sim  500$ arcmin$^{-2}$ for the continuum survey (both assuming 4 hour SKA integrations). For the \HI survey each galaxy is detected at $> 10 \sigma$ to avoid false identification, thus each \HI galaxy has sufficient signal/noise to apply the velocity technique detailed here.

\subsection{Velocity map sensitivity}

A full simulation of the sensitivity of the velocity map technique, incorporating all the instrumental and systematic effects that will be important, is well beyond the scope of this paper. However, we can construct a illustrative toy model.

In our model we use a $10\times 10$ pixel map, with a measurement of the \HI velocity in each pixel. The velocity uncertainty is uniform across the map and no intensity information is used (no isophotes, which should further constrain parameters). For simplicity a single velocity vs. radius template is used (based on the Milky Way), and we allow the amplitude of the velocity and the radial scale to both vary. Our toy model then has 5 parameters: the inclination and orientation of the galaxy spin axis (both prior to lensing), the amplitude and radial scale of the velocity, and the shear. Several hundred realizations are then created, using the significance of the galaxy detection to determine the velocity uncertainty in each pixel.

All five of the parameters are well fit, and normally distributed around the expected values. For a single $10\sigma$ galaxy image, the standard deviation of the shear is 0.02. As expected, the uncertainty in the shear decreases linearly with the brightness of the galaxy, decreasing to 0.007 for a $30\sigma$ galaxy, and 0.002 for a $100\sigma$ detection. Using the galaxy counts and luminosity function of the proposed SKA \HI survey, each square arcminute would have of order 10 galaxies, several of which would be significantly brighter than $10 \sigma$ \citep{AbdallaRawlingsSKA}.

These numbers can be compared to the shear uncertainty of $\sim 0.015$ per arcmin$^{-2}$ in the proposed SKA continuum survey after using 500 galaxy images to beat down the ellipticity noise. While the galaxy counts for the velocity technique are much lower, this toy model suggests that the shear measurements could equal or exceed the continuum observations. 

However, we do not want to over interpret the sensitivity estimates from this toy model. Realistic values for the baseline distribution (PSF), degeneracies due to fitting velocity profiles, and systematics related to non-circular velocity distributions and warped galactic disks all need to be incorporated, preferably based on real \HI observations. Whether \HI or continuum lensing surveys are more sensitive will likely depend on these details of the observational design and systematics.

In addition, different applications may favor different techniques. For mapping cluster mass distributions, very deep \HI observations could map to much smaller mass scales due to the lack of ellipticity noise. While cosmology surveys are simply concerned with the relative signal-to-noise over large volumes. 

It should also be noted that most of the next generation radio telescopes, such as the xNTD and SKA, will use FX correlators. One feature of FX correlators is that the frequency resolution is independent of the bandwidth, so the \HI and continuum surveys can be performed simultaneously with no sensitivity penalty.

\section{Conclusion}

In this paper we have introduced a technique for measuring weak lensing with galaxy velocity maps. The primary advantage of the technique is that it avoids the ellipticity noise inherent in continuum weak lensing surveys. We have also argued that due to the increased utility of bright galaxies, the velocity lensing technique with planned \HI surveys may be competitive with proposed continuum lensing surveys. 

Much additional work is needed to evaluate the real-world utility of the velocity mapping technique. In the future we hope to use VLA and EVLA observations to study the systematic limits in real galaxy observations. Combined with careful studies of the \HI luminosity function and instrumental effects of the xNTD and SKA designs, we hope determine the applicability of this technique to future weak lensing surveys.


\section{Acknowledgements}
I'd like to thank Adam Lidz, T.J. Cox, and Matias Zaldarriaga for very helpful discussions and comments during the writing of this paper.


\begin{thebibliography}{}

\bibitem[Abdalla \& Rawlings(2005)]{AbdallaRawlingsSKA} Abdalla, F.~B., \& 
Rawlings, S.\ 2005, \mnras, 360, 27 

\bibitem[Bartelmann \& Schneider(2001)]{BartelmannSchneiderWL}Bartelmann, M. and Schneider, P., 2001, Phys. Reports, 340, 291--472

\bibitem[Blain(2002)]{Blain2002} Blain, A.~W.\ 2002, \apjl, 570, L51

\bibitem[Blake et al.(2004)]{BlakeSKA} Blake, C.~A., Abdalla, F.~B., Bridle, S.~L., \& Rawlings, S.\ 2004, New Astronomy Review, 48, 1063  

\bibitem[Cabella \& Kamionkowski(2005)]{KamshearNotes2005}Cabella, P., and Kamionkoski, M., 2005, astro-ph/0403392

\bibitem[Kaiser(1998)]{KaiserWL} Kaiser, N.\ 1998, \apj, 498, 26 

\bibitem[Miralda-Escude(1991)]{Miralda-Escude1991} Miralda-Escude, J.\ 
1991, \apj, 370, 1 

\bibitem[Zwaan et al.(2003)]{ZwaanHIPASS} Zwaan, M.~A., et al.\ 
2003, \aj, 125, 2842 

\bibitem[Dark Energy Task Force(2006)]{DETF}Dark Energy Task Force, 2006 ...

\end{thebibliography}
\end{document}